\documentclass[prl,superscriptaddress,twocolumn]{revtex4}
\usepackage{epsfig,amsbsy,amsmath,amsfonts,graphics,latexsym}
\usepackage{eepic,bm,epsfig,color,subfigure}
\def\bra#1{\mathinner{\langle{#1}|}}
\def\ket#1{\mathinner{|{#1}\rangle}}

\newcommand{\pj}[1]{\ket{#1}\bra{#1}}
\newcommand{\beq}{\begin{equation}}
\newcommand{\eeq}{\end{equation}}
\newcommand {\cali}[1]{{\mathcal #1}}
 
\newcommand{\ad}{\text{ad}\,}
\newcommand{\mbf}[1]{\mathbf{#1}}

\newcommand{\Del}{\mbf{\Delta}}
\newcommand{\tr}{{\tt Tr}} 
\newcommand{\hconj}[1]{{#1}^{\dagger}} 
\newcommand{\tp}[1]{{\mathbf #1}} 
\newcommand{\pmat}{\begin{pmatrix}} 
\newcommand{\emat}{\end{pmatrix}} 
\newcommand{\complex}{\mathbb{C}}
{}{}
{}{}
{}{}
{}{}
{}{}

\newcommand{\commentout}[1]{}
\begin{document}
\title{Decoherence-free quantum information in Markovian systems} 
\author{Manas K. Patra} 
\email{manas@cs.york.ac.uk}
\affiliation{Department of Computer Science, University of York, Heslington, York, United Kingdom YO10 5DD}
\author{Peter G. Brooke}
\affiliation{Centre for Quantum Computer Technology and Department of
  Physics, Macquarie University, Sydney, New South Wales 2109, Australia}
\begin{abstract}
Decoherence in Markovian systems can result indirectly from the action of a 
system Hamiltonian
 which is usually fixed and unavoidable.  Here, we
show that in general in Markovian systems, because of the system Hamiltonian,
quantum information decoheres.  We give conditions for the system
Hamiltonian that must be satisfied if coherence is to be preserved.  Finally,
we show how to construct robust subspaces for quantum information
processing.
\end{abstract}
\maketitle
Decoherence remains the most important obstacle to experimental realizations
of quantum processors.  One well-developed method of counteracting the effects of
decoherence is to encode 
quantum information (QI) into decoherence-free subspaces and subsystems
(DFSs)~\cite{Pal96,Duan97,Duan98,Zan97a,Lidar98,Knill00,Lidar03}.
This form of passive error correction has been well-studied, and has led to
(approximate) realizations of DFSs.  For example, in ion traps
Kielpinski \emph{et al}.~used a 
decoherence-free (DF) state of two trapped ions to enable encoded information
to be stored longer than its unencoded counterpart~\cite{Kiel01}, and in an
optical system Kwiat \emph{et al}.~prepared a similar DF state using parametric
down-conversion~\cite{Kw00}.  These 
proof-of-principle experiments have shown that encoding QI into DFSs improves
storage lifetimes, and have partly justified the extensive theoretical
investigations into DFSs.   

The (strict) requirement for infinite-lifetime (DF) quantum information
storage is that all qubits must be symmetrically coupled to the
environment~\cite{Pal96,Duan97,Duan98,Zan97a,Lidar98,Knill00, 
Lidar03}.  Most theoretical results 
regarding DFSs rely on this assumption, which is possible to obtain
only for qubits that are colocated. As the number of qubits increases this approximation becomes
less tenable.  Furthermore, it has recently been shown that
infinite-lifetime quantum information storage is not possible if the physical
qubits of interest are not colocated~\cite{kar07}.  
So, at best we are left with the regime in which all physical qubits {\it
  approximately} experience the same Markovian environment--the qubits
are very close together, but not colocated.  A full characterization of
decoherence in this regime is the purpose of this paper. 

So, we examine the hitherto neglected case of decoherence in physical systems
for which the exact symmetric coupling between the qubits is perturbatively
broken.  One physical example of this is given by closely spaced
dipole-coupled qubits that approximately satisfy the requirements for Dicke
superradiance~\cite{Brooke07b}.  For a full analysis of this
experimentally accessible regime within the Lindblad master equation, the
effects of both the Lamb-shift-type Hamiltonian and the decoherence operators
must be included.  The unitary evolution generated by the Lamb-shift-type
Hamiltonian can cause DFQI to evolve into non-DF states, and so decay via the
action of the decoherence operators.  
Another example is a Heisenberg-type interaction in a spin lattice.  We make no assumptions with regards to the Lamb-shift-type Hamiltonian, and find that for Markovian systems stable quantum information is rare.  We relax the requirement for infinite-lifetime information storage, and derive 
expressions for fidelities which depend on the relative strengths of the evolution operators.
Finally, we show how to construct robust subspaces for quantum information processing.

For the system density-matrix $\rho$ in the Hilbert
space $\cali{H}$, the most general description of Markovian dynamics for
initial decoupling between the system and the bath is given by the Lindblad
master equation
\begin{subequations}\label{eq:Lindblad}
\begin{align}
\dot{\rho}& = -i[\tp{H},\rho]+ L_D(\rho) \\ 
L_D(\rho)&= \frac{1}{2}\sum a_{ij} ([\tp{S}_i,\rho
  \hconj{\tp{S}_j}]+[\tp{S}_i\rho, \hconj{\tp{S}_j}]) \label{eq:1b}, 
\end{align}
\end{subequations}
where $(a_{ij})$ is a time-independent Hermitian coefficient matrix, and
$L_D$ is a completely positive map. The
presence of the decoherence operators $\tp{S}_i$ in Eq.~\eqref{eq:1b} means that the evolution
may not be unitary. But, if $L_D(\rho(t))=0$ then
$\rho(t)   =  e^{-i\tp{H}t}\rho(0)e^{i\tp{H}t}$, so one defines DF dynamics as
satisfying $L_D(\rho)=0$.  One can restrict to pure states 
$\rho=\ket{\psi}\bra{\psi}$~\cite{Lidar98} to give a sufficient condition
for $L_D(\rho) = 0$ as $\tp{S}_i \ket{\psi} =c_i\ket{\psi}$.    
Thus, a DFS is defined as an invariant subspace $M\subset \cali{H}$
such that $\tp{S}_i\ket{\psi}=c_i\ket{\psi}$   
$\forall$ $\psi\in M$.  This condition is guaranteed if $M$ is an
eigenspace of $\tp{S}_i$ with eigenvalue zero, which in many 
cases of interest is true for all generators. We note that in two important
cases this is the only possibility: i) if the Lie algebra generated by
$\tp{S}_i$ is semisimple and ii) if all the generators satisfy $\tp{S}_i^n=0$
for some $n$ (are nilpotent).  More precisely, a subspace $A \in \cali{H}$ is DF if $\tp{S}_i(A)=0$ $\forall$ $i$.
Although, this condition seems somewhat 
stronger than the usual condition of a {\em common} invariant eigenspace with eigenvalue $c_i$ (not necessarily zero) 
of $\tp{S}_i$ the analysis below can be generalized by transforming the operators $\tp{S}_i\rightarrow \tp{S}_i-c_iI$. 
The evolution equation~\eqref{eq:Lindblad} has a unitary and nonunitary part, so even if
$\rho(t_0)$ satisfies $L_D(\rho(t_0))=0$ it does not ensure that
$L_D(\rho(t))=0, \text{ for } t>t_0$.  This motivates the following definition.
A DFS $A$ of $\cali{H}$ is to be considered completely-decoherence-free (CDF) if for
any state $\ket{\alpha}\in A$, $\rho(0)=\ket{\alpha}\bra{\alpha}$ gives
$L_D(\rho(t))=0$.  
This condition is weaker than that derived in
Ref.~\cite{Shab05}, but stronger than that derived in Ref.~\cite{Lidar98}.

We illustrate the regime of interest to this paper using the example of dipole-coupled qubits.  For these qubits, there exists a regime for which   \(\tp{H}= \sum_{ij} \Delta_{ij} \tp{S}^i_+\tp{S}^j_-\).  This occurs when the qubits are closely spaced~\cite{Brooke07b}.  It happens that for closely-spaced dipole-coupled qubits the spatially-dependent interatomic spontaneous emission described by the matrix $( a_{ij})$ no longer depends on index $i,j$, but the spatially-dependent interatomic coherent dipole-dipole interaction does.   This means that there exists exact DF states that are acted upon by $\tp{H}$, causing transitions to non-DF states.  This is the case for any separation greater than zero.

We begin with a criteria for CDF dynamics.

{\bf Proposition 1} Let $V$ be a DFS in $\cali{H}$: $\tp{S_i}\cdot V=0$ $\forall$ $i$. A 
necessary and sufficient condition that $V$ contain a CDFS $W$ is
$\tp{H} \cdot W \subset W$. In particular, $\tp{H}$ can be
diagonalized in $W$.   

{\bf Proof} Define $\rho^\prime(t) = e^{i \mathbf{H} t} \rho(t) e^{-i \mathbf{H} t}$. 
The equation satisfied by $\rho'$ is \( 
\dot{\rho'} = \text{L}^\prime_{\text{D}}[\rho^\prime]\), where $
L'_{D}[\mathbf{S}_i']=L_D 
[e^{i \mathbf{H}t}\mathbf{S}_i e^{-i\mathbf{H}t}]$. Hence,
in this picture $\rho^\prime(0)$ is DF iff
$ L'_D [\rho^\prime(0)]=0$.  The generic
DFSs are spanned by vectors $\ket{x}$ such that
\(\mathbf{S}_i'\ket{x}=e^{i\mathbf{H}t} 
\mathbf{S}_ie^{-i\mathbf{H}t}\ket{x}=0\). Let $W$ be the subspace consisting
of all such vectors.  This must be satisfied for all $t$, so we
have $\mathbf{H} \cdot W \subset W \;\forall i$.  Conversely, if this is
satisfied then  
\(\mathbf{S}_ie^{-i\mathbf{H} t}\ket{x}=0 \;\forall i\).

A useful consequence of the proposition is the following. 

{\bf Corollary} A subspace $W$ is CDF iff $(\ad(\tp{H}))^n(\tp{S}_i)\cdot
W=0,\; \forall n\text{ and }i$. This is equivalent to the condition
$[\tp{H}^n,\tp{S}_i]\cdot W=0$.  

{\bf Proof} The first condition follows from the proposition and the identity
\(e^{it\ad\tp{H}}(\tp{S}_i)=e^{it\tp{H}}\tp{S}_ie^{-it\tp{H}}\). The second
condition is proved by induction~\cite{Humphreys}. 

The Pauli matrices $\{\sigma_x,\sigma_y,\sigma_z\}$
generate the Lie algebra 
$su(2)$ and form a basis (with identity matrix $I$) for the space of
observables for each qubit. Let \(\tp{S}^i_a= I\otimes \cdots \otimes
\sigma_a\otimes I\cdots \otimes I,\; a\in \{x,y,z\}\), with $\sigma_a$ only at
the $i^{\text{th}}$ place. Let \(\tp{S}^i_{\pm}=
\tp{S}^i_x\pm i\tp{S}^i_y$, $\tp{S}_{\pm}= \sum_i \tp{S}^i_{\pm}\) and
$\tp{S}_z=2\sum_i\tp{S}^i_z$. These operators define a representation of
$su(2)$ on the system Hilbert space $\cali{H}= \otimes 
^{N}\mathbb{C}^2=\mathbb{C}^{2N}$, and satisfy \([
  \tp{S}_{\pm}, \tp{S}_z]= -2\tp{S}_z \text{ and }
[\tp{S}_+,\tp{S}_-]=\tp{S}_z\). We denote a tensor basis for
$\complex^{2N}$ as $\ket{i_1}\otimes  \cdots \ket{i_N}$ where
$i_k\in \{0,1\}=\ket{i_1\cdots i_N}$, and we write $\ket{j_1\cdots j_k}$ for
a vector with ones at $j_1\cdots j_k $ and zeros elsewhere. The Hamiltonian
is written \(\tp{H}= \sum_{ij} 
\Delta_{ij} \tp{S}^i_+\tp{S}^j_-\), where $(\Delta_{ij})$
is a Hermitian matrix.  Without loss of
generality, we take $\mbf{\Delta}$, for 
$\mbf{\Delta}=(\Delta_{ij})$, to be real symmetric. We consider the case
where $a_{ij} = a$, so there is only one 
Lindblad generator $\tp{S}_-$.    In this instance, Eq.~\eqref{eq:1b} takes the form
$L_D(\rho) = \kappa(\tp{S}_-\rho\tp{S}_+
-\frac{1}{2}(\tp{S}_+\tp{S}_-\rho+\rho\tp{S}_+\tp{S}_-))$.  Each irreducible 
representation (irrep) of $\cali{H}$ is generated by a unique lowest weight
vector with weight $-(r-1)$ for irrep dimension $r$ that 
satisfies $\tp{S}_-\ket{\alpha}=0$.    Note that $\tp{H}$ leaves the weight
spaces invariant, so we only consider DFSs of fixed weight.

The subspace $V_1$ is 
generated by the basis \( B\equiv \{\ket{i} 
=\ket{0\cdots 010\cdots 0} \; | \text{ for 1 in the $i^{\text{th}}$
  place}\}\).  Here,
we let $\tp{H}$ also stand for its restriction to $V_1$. Then, $\tp{H}\ket{i}
= \sum_k \Delta_{ki}\ket{k}$. A state 
$\ket{\mbf{x}} = \sum_i x_i \ket{i}$ is a lowest weight state
iff $\sum_i a_i=0$. There are $N-1$ such independent vectors which generate a
DFS, called here $D_1$.  We wish to find out whether there are
any CDFS $\neq 0$ inside $D_1$.  Note that we refer to
{\it the} CDFS $C$ of $D_1$, which denotes the maximal subspace that is the
sum of all CDFSs.  The condition for CDF dynamics is $\tp{H}\cdot C\subset C$.
So, for a nonzero 
subspace $C$ to exist it is necessary and sufficient that $\tp{H}$ have an
eigenvector in $V_1$. In the fixed basis $B$, we represent an arbitrary
$\ket{\mbf{x}} = \sum_i x_i \ket{i}$ as a column vector $\mbf{x}^T = (x_1,
\cdots, x_N)$ in $\mathbb{C}^N$. It is clear that $\ket{\mbf{x}}$ is an
eigenvector of $\tp{H}$ iff $\mbf{x}$ is an eigenvector of $\Del$ 
with the same eigenvalue. Hence, there will be a nonzero CDFS iff $\Del$
has an eigenvector $\mbf{x}$ such that $\tr(\mbf{x})=0$ where $\tr(\mbf{x})=
\sum_i^N x_i$.  Suppose $\Del$ has a degenerate eigenvalue $c$. Then
there are at least two independent vectors $\mbf{y}\text{ and } \mbf{z}$. If
we have $\tr(\mbf{y})=\tr(\mbf{z})=0$, then there is a nonzero CDFS
containing at least $\ket{\mbf{y}} \text{ and } \ket{\mbf{z}}$. Otherwise,
suppose $\tr(\mbf{z})=k\neq 0$. Then, $\mbf{x}\equiv \mbf{y}
-\frac{\tr(\mbf{y})}{k}\mbf{z}$ has trace zero and we have a nontrivial
CDFS. This gives:

{\bf Proposition 2}  A sufficient condition for the existence of nonzero CDFS is that the matrix
$\Del$ has a degenerate eigenvalue $c$. If $c$ is $m$-fold degenerate, then
the dimension of the CDFS $\geq m-1$. 

Although the main result of this paper concerns the rarity of CDFSs, there are two examples of physical systems that satisfy CDF dynamics.  First, consider four qubits in a spatially symmetric configuration, e.g., the corners of
a square lattice. Then, from symmetry considerations it is clear that the row (or column)
sums of the matrix $\Delta$ is constant. This automatically guarantees the condition in the 
proposition. There is at least one CDFS containing $\sum_{i=1}^4 \ket{i}$.  Second, consider two dipole-coupled two-level atoms.  In the Dicke limit, the single-excitation antisymmetric state is CDF because for the special case of two atoms, the dipole-dipole interaction does not cause information to leak from the DF state to the non-DF state.    
Consider now the general case for a single-excitation. Assume all the
eigenvalues of $\Del$ are non-degenerate.  We seek a condition on $\Del$ that
will ensure the existence of an eigenvector in $D_1$. Suppose
$\mbf{x}=(x_1,\cdots,x_N)^T$ is such an eigenvector with eigenvalue $c$. Then
the corresponding eigenspace is 1-dimensional, and we have the following set
of equations 
$ \label{eq:basicFirstEx1}
\Delta_{11}x_1+ \cdots +\Delta_{1N}x_N  = cx_1$;
$\Delta_{N1}x_1+ \cdots +\Delta_{NN}x_N  = cx_N$ ; 
$x_1+ \cdots +x_N  = 0$.
If we set one of the components, say $x_N=1$, these equations have a
unique solution.  We rewrite the first $N$ 
equations as 
$(\Delta_{11}-c)x_1+ \cdots +\Delta_{1,N-1}x_{N-1}  = - \Delta_{1N}$;
$\Delta_{N-1,1}x_1+ \cdots +(\Delta_{N-1,N-1}-c)x_{N-1} = -\Delta_{N-1,N}$;
$\Delta_{N1}x_1+ \cdots +\Delta_{NN} -c = 0$.
Let $\Gamma(c)$ denote the $(N-1)\times(N-1)$ matrix such that
$\Gamma_{ij}=\Delta_{ij}- c\delta_{ij},\;1\leq i,j \leq N-1$ (for
$\delta_{ij}$ the Kronecker $\delta$). The uniqueness of $\mbf{x}'\equiv
(x_1,\dotsc, x_{N-1})^T$ 
implies that $\Gamma$ is invertible. Writing $\mbf{d}= -(\Delta_{1N},\dotsc,
\Delta_{N-1,N})^T$ we have $\mbf{x}' = \Gamma^{-1}(c)\mbf{d}$. The last
equation  $\Delta_{N1}x_1+ \cdots +\Delta_{NN} -c = 0 $ can be written as
$\mbf{d}^T\mbf{x}'-c+\Delta_{NN}$. Hence, we have
$\mbf{d}^T\Gamma^{-1}(c)\mbf{d}-c+\Delta_{NN}=0$. That is,  
$f(c)=\mbf{d}^T\text{adj}(\Gamma(c))\mbf{d}-\det{(\Gamma(c))}(c-\Delta_{NN})=0$
,  where $\text{adj}(\Gamma)$ is such that $\text{adj}(\Gamma)\Gamma=
\det{(\Gamma)}I$. From the condition $\tr(\mbf{x})=0$ and $x_N=1$ we obtain
$\tr(\Gamma^{-1}(c)\mbf{d})=-1$. Hence,
$g(c)=\tr(\text{adj}(\Gamma(c))\mbf{d})+\det{(\Gamma(c))}=0$.  So, we get two
polynomial equations in $c$ whose coefficients are functions of 
$\Delta_{ij}$. For a solution to exist the resultant of the two polynomials
must vanish~\cite{Waerden}, and we get a polynomial relation
$R(\Delta_{ij})$ among the $\Delta_{ij}$ which does not vanish
identically. There are $N(N+1)/2=M$ independent parameters
characterizing any real symmetric matrix. Hence, the space of all such
matrices
may be identified with ${\mathbb R}^M$.  We have just seen that
for $\Del$ to have a nontrivial CDFS, it must satisfy (at least)
one polynomial equation. So, there are Hamiltonians with matrix $\Del$ which
{\em do not} have any CDFS in $V_1$.  We will see that this is the norm rather that the
exception---Hamiltonians with CDFS are rare.

We illustrate the above analysis with an example.  The first excited subspace
$V_1$ is spanned by $\{ \ket{1}=\ket{001},\ket{2}= \ket{010},
\ket{3}=\ket{100}\}$. Let the matrix $\Delta$ corresponding to $\tp{H}$
restricted to $V_1$ be given by    
\(\Delta_{12}=\Delta_{21}=x_3, \Delta_{13}=\Delta_{31}=x_2 \text{ and }
\Delta_{23}=\Delta_{32}=x_1\) with diagonal elements zero. The CDFS condition
translates to $x_3=x_1 \text{ or } x_2$ corresponding to 
eigenvalues $-x_2$ and $-x_1$ respectively. The third possibility $x_1= x_2$
gives eigenvalue $-x_3$.  Fixing 
$x_3=x_1=a$ the eigenvalues of $\Delta$ are 1 and 
$x_2\pm\sqrt{x_2^2+8a^2}$. The only possibility for a degenerate eigenvalue is
when $x_1=x_2=x_3$.  We conclude that $\tp{H}$
restricted to the first excited subspace $V_1$ will 
  have an eigenvector in $D_1$ (DF subspace) iff at least two of its off-diagonal
  entries are equal, and it will have two eigenvectors in $D_1$ if all
  three are equal. The parameter space of $\tp{H}$ can be identified
with $\mathbb{R}^3$ as $x_1,x_2\text{ and } x_3$ the range over the real
numbers. Then, the only Hamiltonians with eigenvectors in $D_1$ are
characterized by the parameters that lie in the planes $x_1=x_2,x_1=x_3
\text{or } x_2=x_3$. So, we conclude that the Hamiltonians which leave
some DFS state in $V_1$ invariant is a negligible fraction of all the
possible Hamiltonians---in general, quantum information will decay.

We know that the DFS in the $m^{\text{th}}$ excited
subspace $V_m$ is spanned by vectors $\ket{\mbf{x}_m}= \sum x_{i_1,\dotsc,
  i_m}\ket{i_1,\dotsc, i_m}$ such that $M'=\binom{N}{m-1}$ equations $
\sum_{i_r=1}^N
x_{i_1,\dotsc,i_r,\dotsc, i_m} = 0\; (*)$---with all indices except $i_r$ 
fixed---are satisfied.  The action of $\tp{H}$ on $V_m$ is more complicated:
\begin{align}
 \tp{H}\ket{\mbf{x}_m}  = \sum_{i_1,\dotsc,i_m}x_{i_1,\dotsc,i_m}( \sum_k
 \Delta_{ki_1} \ket{ki_2i_3\cdots i_m}+  \nonumber \\
\cdots+\sum_k \Delta_{ki_m}\ket{i_1i_2i_3\cdots k}). 
\end{align}
We require that $\ket{\mbf{x}_m}$ satisfy
$\tp{H}\ket{\mbf{x}_m}=\lambda\ket{\mbf{x}_m}$.    
We write the matrix representing $\tp{H}$ restricted to the subspace $V_m$
as $\Delta^{(m)}$. The action of $\tp{H}$ on $V_m$ is equivalent to
that of $\Delta^{(m)}$ on $\mathbb{R}^{M}$ for $M= \binom{N}{m}$
whose coordinates are given by $x_{i_1,\dotsc,i_r,\dotsc, i_m}$.  We write the
eigenvalue equations as in the previous section. Since the eigenvector must
satisfy $M'$ equations $(*)$ we use
them to write the last $M'$ 
components of such a vector in terms of the first $M-M'$, and substitute in the
eigenvalue equation of $\Delta^{(m)}$. The resulting system of equations in
$M-M'$ variables must have rank less than $M-M'$ for a nontrivial
solution to exist. Let $\Delta'^{(m)}$ be the square matrix of the coefficients
of the first $M-M'$ equations. It must have determinant zero. This gives a
polynomial equation in $\lambda$, the eigenvalue.  Write
$f(\lambda)=\text{det}(\Delta'^{(m)}(\lambda))$, which shows the explicit
dependence on the eigenvalue $\lambda$.  We also have $g(\lambda)=
\text{det}(\Delta^{(m)}(\lambda))=0$, the original characteristic equation. The
coefficients in $f(\lambda)$ and $g(\lambda)$ are functions of the 
variables $\Delta_{ij}$. The necessary and
sufficient condition that $\tp{H}$ has an eigenvector in a DFS is that
$f(\lambda)$ and $g(\lambda)$ have a common root: that is, the resultant
$R(\Delta)=0$. It can be shown that the resultant does not vanish
identically. All possible Hamiltonians, parametrized by the real numbers
$\Delta_{ij}$, constitute a 
manifold of dimension $N(N+1)/2$. The Hamiltonians which have an eigenvector
in a DFS in $V_m$ lie on a submanifold of dimension strictly less than
$N(N+1)/2$. Hence, using Sard's theorem~\cite{Milnor65} and 
generalizing to Hermitian matrices we arrive at the
following theorem. 

{\bf Theorem 1} Let $X$ be the $N^2$-dimensional real manifold of the parameters
characterizing the possible Hamiltonians in the Lindblad master equation. Let
$S$ be the set of values of the parameters which characterize Hamiltonians
that have at least one DF state other than the ground state evolving into a
DFS state at all times. Then $S$ is of measure zero in $X$.  

{\bf Proof} Let $S_m$ be those members $S$ that correspond to Hamiltonians that have some
DF states in a fixed weight space $V_m,\; m>0$ evolving into DF states at all
times. From Proposition 1, we see that this is equivalent
to the condition that the Hamiltonians in $S_m$ have an eigenvector in the
DFS in $V_m$. From the preceding discussion, $S_m$ has measure zero in
$X$. Since $S=\cup_m S_m$, $S$ too has measure zero.  

Informally, we can say that for almost all Hamiltonians there is no DF state
other than the ground state which evolves into other DF states at all
times. Therefore, the best we can hope for is to seek states 
which remain DF up to some orders of perturbation. 

Corollary to Proposition 1 implies
that to get DF states 
we should look for states that are annihilated by operators
$\tp{S}_-,[\tp{H},\tp{S}_-],[\tp{H}[\tp{H},\tp{S}_-]],\dotsc$.  
We have seen that in general DFSs are not invariant under
$\tp{H}$.  So, we seek invariance up to certain orders. The condition
for states $\ket{\alpha}$ that are invariant  
up to first order (in $\tp{H}$) is that $\tp{S}_-\ket{\alpha}=[\tp{H},\tp{S}_-]
\ket{\alpha}=0$. Similarly, the second order condition is
\(\tp{S}_-\ket{\alpha}=[\tp{H},\tp{S}_-] \ket{\alpha}=
     [\tp{H},[\tp{H},\tp{S}_-]]\ket{\alpha}=0\).  We assume that the
     initial state $\rho(0)=\ket{\psi}\bra{\psi}$ is a DF state. Then
     $\rho(t)= e^{-it(\ad \tp{H}+i L_D)}\rho(0)$ and if the evolution is
     unitary, $\rho'(t) = e^{-it\tp{H}}
     \ket{\psi}\bra{\psi}e^{it\tp{H}}$. Hence, we take   
$F^2(\rho'(t),\rho(t))=\bra{\psi}e^{it\tp{H}}e^{-it(\ad{\tp{H}}+i
  L_D)}\rho(0)e^{-it\tp{H}}\ket{\psi} 
$ 
as a measure of deviation from unitary evolution~\cite{Jozsa94}.  Writing
$\ad{\tp{H}}=X$, we 
have \(F^2= \bra{\psi}e^{itX}e^{-it(X+iL_D)}\ket{\psi}\).  Using the
Zassenhaus formula~\cite{Wilcox67}, 
the (exact) fidelity can be written 
\(
F^2=  \bra{\psi}e^{L_D}e^{-it^2[X,L_D]/2} 
 e^{-t^3Z_3(X,L_D)} 
\cdots \rho(0)\ket{\psi}.
\) where 
$Z_3(X,L_D)=[X,[X,L_D]]/6+i[L_D,[L_D,X]]/3)$
We consider the following three approximate cases. 

{\em Case 1 (weak unitary part)}. The Hamiltonian $\tp{H}$ is replaced by
$\epsilon\tp{H}$, and treated as a perturbation. Then up to first order in
$\epsilon$, \(F^2= 
\bra{\psi}e^{tL_D}e^{-it^2[L_D,X]/2}\cdots
e^{c_kt^{k}(\ad{L_D})^{k-1}(X)}\rho(0) \ket{\psi}\) where $c_k= (-1)^{k-1}
\epsilon(k-1)i/k!,\; k >1$.  The state $\rho(0)=\pj{\psi}$ is DF, so 
$F^2=\bra{\psi}(1+\sum c_k t^k 
(\ad{L_D})^{k-1}(X))\rho(0)\ket{\psi}$. 
For DF-states \(\bra{\psi}(\ad{L_D})^{k-1}(X)\rho(0)\ket{\psi}=0\) and 
$F^2=1$~\cite{Bacon99,Lidar98}.   

{\em Case 2 (strong unitary part)}. The Lindblad part is treated as a
perturbation to give $\dot{\rho}=(-i\ad{\tp{H}}+\epsilon L_D)\rho$. Then up to
first order in $\epsilon$, \(F^2= \bra{\psi}e^{tL_D}e^{it^2[X,L_D]/2}\cdots
e^{d_kt^{k}(\ad{X})^{k-1}(L)}\rho(0) 
\ket{\psi}\) where $d_k= (-i)^{k}\epsilon/k!,\; k >1$.  Computing the series
expansions to first order, we get $F^2= \bra{\psi}(1+\sum d_kt^k
(\ad{X})^{k-1}(L_D))\rho(0)\ket{\psi}$. Unlike the previous case
$L_D(\rho(0))=0$ does not guarantee $F^2 = 1$ for all times. There are two
possible courses of action. First, if the time scales 
of $\tp{H}$ are much shorter than $L_D$, zeroth order may suffice. Thus,
instead of talking the initial state $\rho(0)$ as an eigenstate of Lindblad
generators, we use an eigenstate of $\tp{H}$. This method is used 
in Ref.~\cite{BCP07}. 

{\em Case 3 (short-time expansions)}.  Note that the 
Zassenhaus formula shows that DFSs are stable up to first order in
$t$. If we require stability up to order $t^k, \; k \geq 1$, we can work in
the smaller 
subspaces $W$ satisfying $\ad{\tp{H}}^m(\tp{S}_i)\cdot W=0,\; m=0,\dotsc, k-1$
$\forall$ $i$. Since we are dealing with finite-dimensional Hilbert spaces,
it suffices to require stability up to order $n$, the dimension of the system
space. However, we have seen that this is not possible in general
(Theorem 1), so we have to be satisfied with smaller $k$.

We consider the universality of
subspaces for external Hamiltonians that implement quantum gates. Let
$\tp{H}_c$ denote the external Hamiltonian and $\tp{H}_d = \sum_{ij} 
\Delta_{ij} \tp{S}^i_+\tp{S}^j_-$. The equation of motion is 
$\dot{\rho} =
(-i(\ad{\tp{H}_c}+\ad{\tp{H}_d})+L_d)\rho\equiv (Y+X+L_D)\rho$. 
Since we aim to work in DFSs, they must be invariant under any
external Hamiltonian. Thus, $[\tp{H}_c,\tp{H}_i]=0$, and $[Y,L_D]=0$. Assuming
that the characteristic time-scale of $\tp{H}_c$ is much 
smaller than other operators (essential for feasible
computation), then using the short time expansion we can construct
robust subspaces for computation. For example, if we wish to terminate at
third order, then a subspace $W$ satisfying $L_D\cdot W=[\tp{H}_d,L_d]\cdot
W=
[\tp{H}_d,[\tp{H}_d,L_D]]\cdot W=0$ will suffice.  

We have shown that in Markovian systems the Lamb-shift-type system 
Hamiltonian generally causes transitions from DF states to non-DF states.
The results presented here emphasize the importance of accounting for {\it
  both} unitary and nonunitary evolution in passive quantum error correction.
Note that the results presented here, as long as the Lindblad operators do not
cause transitions between irreps, can be extended to the finite
temperature case.   
\bibliography{dfstheory2}    
\end{document}